\documentstyle[12pt]{article}
\if@twoside\oddsidemargin25mm
\else\oddsidemargin25mm\evensidemargin25mm\marginparwidth25mm\fi%
\batchmode
\begin{document}
\newcommand{\half}{{\textstyle\frac{1}{2}}}
\newcommand{\ft}[2]{{\textstyle\frac{#1}{#2}}}
\newcommand{\cmap}{{\bf c} map}
\newcommand{\rmap}{{\bf r} map}
\newcommand{\crmap}{{{\bf c}$\scriptstyle\circ${\bf r} map}}
\newcommand{\Ka}{K\"ahler}
\newcommand{\Al}{Alekseevski\v{\i}}
\newcommand{\eqn}[1]{(\ref{#1})}
\newcommand{\QED}{{\hspace*{\fill}\rule{2mm}{2mm}\linebreak}}
\newcommand{\Bk}{B_1\cdots B_k}
\newcommand{\Bpmt}{B_1\cdots B_{p-2}}
\begin{titlepage}
\begin{flushright} CERN-TH.6593/92\\ KUL-TF-92/30\\ THU-92/17\\
hepth@xxx/9207091
\end{flushright}
\vfill
\begin{center}
{\large\bf Broken sigma-model isometries in very special geometry} \\
\vfill
{\bf Bernard de Wit$^1$}\\
Theory Division, CERN\\ CH-1211 Geneva 23, Switzerland\\[0.3cm]
and \\[0.3cm]
 {\bf Antoine Van Proeyen$^2$} \\
Instituut voor theoretische fysica
\\Universiteit Leuven,
B-3001 Leuven, Belgium
\end{center}
\vfill
\begin{center}
{\bf ABSTRACT}
\end{center}
\begin{quote}
We show that the isometries of the manifold of scalars in
$N=2$ supergravity in $d=5$ space-time dimensions can be broken by
the supergravity interactions. The opposite conclusion holds for the
dimensionally reduced $d=4$ theories, where the isometries of the
scalar manifold are always symmetries of the full theory. These
spaces, which form a subclass of the {\em special} K\"ahler
manifolds, are relevant for superstring compactifications.
\vskip 2mm \hrule width 5.cm \vskip 1.mm
{\small\small
\noindent $^1$ On leave of absence from the Institute for
Theoretical Physics, Utrecht, The Netherlands\\
\noindent $^2$ Onderzoeksleider, N.F.W.O. Belgium;
Bitnet FGBDA19 at BLEKUL11}
\normalsize
\end{quote}
\begin{flushleft} CERN-TH.6593/92 \\ KUL-TF-92/30\\ THU-92/17  \\
July 1992
\end{flushleft}
\end{titlepage}

We have grown accustomed to supersymmetry implying the existence
of symmetries rather than breaking them. In this letter we
approach this question from the other end and investigate
what happens to the symmetries of
certain non-linear sigma models when contained
in $N=2$ supergravity. We analyze two supergravity theories
coupled to vector supermultiplets, both based on the real symmetric
tensors
$d_{ABC}$ ($A,B,\ldots = 1, \ldots, n$). One is supergravity in five
space-time dimensions \cite{GuSiTo}; the other one is the
same theory but reduced to four space-time dimensions \cite{dWVP}.

In four dimensions the
supergravity theories based on the tensor $d_{ABC}$ can lead to flat
potentials \cite{BEC}. Furthermore these theories appear
as low-energy effective theories of certain
superstring compactifications on (2,2) superconformal theories
(containing the Calabi-Yau three-folds),
where the $d_{ABC}$ correspond to the Yukawa couplings and
depend on the moduli of the superconformal theory
\cite{CecFerGir,DixKapLou}. The corresponding
non-linear sigma models (as well as
the moduli space of the superconformal theories) exhibit
{\em special} geometry \cite{special}.

In the five-dimensional case we show that the full isometry
group of the corresponding non-linear sigma model is not
necessarily preserved by the supergravity interactions. This
will be done in the context of a specific class of sigma models,
namely those corresponding to an $SO(n-1,1)/SO(n-1)$ target space. We
find that not all the $SO(n-1,1)$ isometries are symmetries of the
full supergravity action. Here we
should stress that most of the attention was focused on the
latter symmetries \cite{GuSiTo}; usually one simply assumes that
those comprise all symmetries of the non-linear sigma model.

This above result is relevant for understanding the symmetry
structure of the non-linear sigma model that emerges after
dimensional reduction, which is governed by the symmetry of the
full action
before dimensional reduction, rather than by the symmetries of
the scalar sector alone, as some of the scalar fields
after dimensional reduction originate from tensor and vector
fields. As was shown in \cite{CecFerGir} dimensional reduction
is a useful tool in understanding the relation between
the low-energy theories corresponding to
different strings compactified on the same superconformal theory.
Its implications for the symmetry structure of the various
geometries is discussed in more detail in \cite{dWVPSwa}.

After dimensional reduction one obtains a non-linear sigma model
whose target space is a special \Ka\ space. Such spaces are
characterized by holomorphic functions that are
homogeneous of second degree. In the case at hand the
relevant functions are
\begin{equation}
F(X)= i d_{ABC} {X^AX^BX^C\over X^0}\ .\label{Fd4}
\end{equation}
Surprisingly enough, in four dimensions
the situation is just the opposite: all isometries of the
corresponding non-linear sigma model constitute symmetries of the
full supergravity action. The proof of this result will be given
in the second part of this letter. The symmetries take the form of the
so-called  duality invariances under which the equations of
motion, but not necessarily the full action, are invariant. For
the class of theories based on \eqn{Fd4} the generalized duality
transformations were derived in \cite{BEC}.
\vspace{7mm}

We start by recalling some relevant features of the Maxwell-Einstein
supergravity theories in $d=5$ space-time dimensions \cite{GuSiTo}.
These theories are based on $n-1$ real scalar fields and $n$ vector
fields (one of them corresponding to the graviphoton). The part of the
Lagrangian pertaining to these fields, reads
\newpage

\begin{eqnarray}
e^{-1}{\cal L} &=& -{\textstyle{1\over 2}} R -{\textstyle{3\over
2}}\,d_{ABC} \,
h^A\,\partial_\mu h^B\,\partial^\mu h^C  \nonumber\\
&& +{\textstyle{1\over 2}}\,\left( 6(d\,h)_{AB} - 9 (d\,h h)_{A}\,
(d\,hh)_B \right) F^A_{\mu\nu}(A) \,F^{B\,\mu\nu}(A)
\nonumber\\
&& + e^{-1} i\epsilon^{\mu\nu\rho\sigma\lambda} d_{ABC} \,
F^A_{\mu\nu}(A)\,F^B_{\rho\sigma}(A)\, A^C_\lambda ,
\label{5dLagr}
\end{eqnarray}
where $(d\,h)_{AB}= d_{ABC}\,h^C$, $(d\,h\,h)_A= d_{abc}\,h^Bh^C$,
etc.. In \eqn{5dLagr} $A^A_\mu$ and $F_{\mu\nu}^A(A)$ denote the
gauge fields and their corresponding Abelian field strengths
while the scalar fields $h^A$ are subject to the condition
\begin{equation}
d_{ABC}\,h^A h^B h^C =1.  \label{dhhh1}
\end{equation}
The scalar fields must be restricted to a domain
so that all kinetic terms in (\ref{5dLagr}) have the required
signature.

The Lagrangian (\ref{5dLagr}) is manifestly invariant under linear
transformations of the fields
\begin{equation}
h^A \to \tilde B^A_{\;B}\, h^B, \qquad  A^A_\mu \to \tilde
B^A_{\;B}\,A^B_\mu,
\end{equation}
provided that the matrices $\tilde B$ leave the tensor $d_{ABC}$
invariant
\begin{equation}
\tilde B^D_{\,(A}\,d_{BC)D}= 0 . \label{Binv}
\end{equation}

However, the corresponding non-linear sigma model may possess
extra invariances, which are not symmetries of the full action.
The first topic of this letter is to establish that this phenomenon
does indeed take place in the context of an explicit class of
models. The corresponding five-dimensional
supergravity theories were already discussed in \cite{GSTkinks}.
The scalar fields parametrize the coset space
\begin{equation}
{SO(n\!-\!1,1)\over SO(n\!-\!1)}\ .\qquad (n\geq 3)
\end{equation}
Assuming that the space remains symmetric in the reduction to
four dimensions, the authors of \cite{GSTkinks} incorrectly
identified the target spaces of the corresponding $d=4$ models
with the hermitian
symmetric spaces ${SU(n,1)/(SU(n)\otimes U(1))}$.
We shall see, however, that some of the symmetries contained in
$SO(n-1,1)$ are
broken by the supergravity interactions, so that the full action
\eqn{5dLagr} is not invariant under this group, and neither is
the dimensionally reduced $d=4$ action. It is therefore not
surprising that the corresponding special \Ka\ manifolds are
no longer symmetric, although they are still homogeneous.
The homogeneity is a consequence of the fact that the solvable
part of $SO(n-1,1)$ is still realized as a symmetry. These homogeneous
spaces were discovered in \cite{dWVP3}, where they were denoted by
$L(-1,n-2)$. By further reduction to $d=3$ dimensions they give
rise to homogeneous quaternionic spaces, which are missing in the
classification of \cite{Aleks}. For a more detailed
discussion of this, we refer to \cite{dWVPSwa}.

Let us now introduce the models. Splitting the range of the indices $A$
into $A=1,2$ and $A=i=3,\ldots,n$, the non-zero $d$ coefficients are
\begin{equation}
d_{122}=1\ ;\qquad d_{2ij}=-\delta _{ij}.  \label{dnew}
\end{equation}
The symmetries of the $d$ tensor consist of $SO(n-2)$ rotations
on $h^i$ and $n-1$ transformations, which are related to $n\times
n$ matrices $\tilde B_{(2)}$ and $\tilde B_{(i)}$, defined by
\begin{eqnarray}
\tilde B_{(2)}{}^1{}_1=\textstyle{\frac{4}{3}} \ ;\qquad
\tilde B_{(2)}{}^2{}_2=-\textstyle{\frac{2}{3}}\ ;\qquad
\tilde B_{(2)}{}^i{}_j=\textstyle{\frac{1}{3}}\delta ^i_j\ ,\nonumber\\
\tilde B_{(i)}{}^j{}_2=\delta ^j_i\ ;\qquad
\tilde B_{(i)}{}^1{}_j=2\delta _{ij}\ \ .\quad{~} \label{tildeBnew}
\end{eqnarray}

Already at this point one can see that something special should
occur. Consider the case $n=3$. The isometries associated with
$\tilde B_{(2)}$ and $\tilde B_{(3)}$ generate a solvable group
that acts transitively on the two-dimensional
space. Hence the space is homogeneous. Spaces that allow
a solvable transitive group of motions are called {\em normal}.
They are isomorphic to the group space of this solvable
group and the corresponding solvable algebra follows from the
Iwasawa decomposition of the full isometry algebra (see, e.g.
\cite{Helg}). As the real space has only two dimensions, the
(non-Abelian) solvable algebra can only be isomorphic to $[t_0,
t_1]=t_1$. But this is just the solvable algebra corresponding to
the symmetric space $SO(2,1)/SO(2)$, which has a
three-dimensional isometry group. This in contrast with the result
found above (cf. \eqn{Binv}), which indicates the existence of
only two symmetries. Therefore we expect that the scalar manifold should
exhibit an extra isometry, which is not a symmetry of
\eqn{5dLagr}. This is indeed confirmed by the following analysis.

Choose $n-1$
independent coordinates $\phi ^a=\{\phi ^2,\phi ^i\}$, such that
\eqn{dhhh1} is satisfied,
\begin{equation}
h^2=\phi ^2\ ;\qquad h^i= \phi ^2\phi ^i\ ;\qquad h^1=
\textstyle{\frac{1}{3}}\left(
\phi ^2\right) ^{-2}+\left( \phi ^2\right) \left(
\phi ^i\right) ^2. \end{equation}
{}From \eqn{5dLagr} we obtain the metric
\begin{equation}
g_{ab}=-d_{ABC}\frac{\partial h^A}{\partial \phi ^a}
\frac{\partial h^B}{\partial \phi ^b}h^C\ ,
\end{equation}
which, in the parametrization above,  takes the form
\begin{equation}
g_{22}=\left( \phi ^2\right) ^{-2} \ ;\qquad g_{2i}=0\
;\qquad
g_{ij}=\left( \phi ^2\right) ^3\delta _{ij}.
\end{equation}

One may now determine all invariances of
$g_{ab}\partial _\mu \phi ^a\partial^\mu \phi ^b$. First there
are $SO(n-2)$ rotations on $\phi ^i$, which constitute an obvious
invariance. Furthermore one finds $2n-3$ additional
symmetries, with parameters $a^i,\,b^i$ and $c$,
\begin{equation}
\delta  \phi ^2 =\textstyle{\frac{2}{3}}b^i\,\phi^i\phi ^1
-\textstyle{\frac{2}{3}}c\,\phi ^2 \ ;\qquad
\delta \phi ^i=b^i\left( \textstyle{\frac{2}{9}}\left( \phi ^2\right)^{-
3}-\half\phi^j\phi ^j\right)  +c\,\phi ^i +a^i.
\label{abcnew}
\end{equation}
When combined with $SO(n-2)$, these variations generate the group
$SO(n-1,1)$.
The transformations parametrized by $c$ and $a^i$ correspond to
$\tilde B_{(2)}$ and $\tilde B_{(i)}$, respectively. However, the
transformations parametrized by $b^i$
are new and are not realized as symmetries of the Lagrangian
\eqn{5dLagr}.

This thus proves the first assertion of this paper according to
which the isometries of the non-linear sigma model can be broken
by supergravity interactions, at least in $d=5$ Maxwell-Einstein
supergravity.
\vspace{7mm}

We now come to the second topic and consider the $d=4$
supergravity theory based on the functions \eqn{Fd4}, where the
tensor $d_{ABC}$ is arbitrary. This theory can be obtained by
dimensional reduction from \eqn{5dLagr}. In this reduction we
obtain complex scalars $z^A$. Their imaginary part originates
from the components of the gauge fields $A^A_\mu$ in the fifth
dimension, while their real part corresponds to the $n-1$
independent fields $h^A$ and the component $g_{55}$
of the metric.
These scalars define a {\em special} \Ka\ space, but of a very special
form \cite{BEC}, namely based on \eqn{Fd4}.
An important property of these models is that, within the
equivalence class of \Ka\ potentials
$K(z,\bar z)\sim K(z,\bar z)+\Lambda (z)+\bar \Lambda (\bar
z)$, there are representatives of the potential that depend only
on the imaginary part of $z^A$, i.e. on $x^A$ defined by
\begin{equation}
z^A\equiv\half (y^A-ix^A)\label{zxy}\ .
\end{equation}
We will denote one such a representative by $g$
\begin{equation}
K(z,\bar z)\sim g( x) = \log (d_{ABC}\,x^Ax^Bx^C) \ .
\end{equation}
Therefore the metric depends only on $x$,
and is just the second derivative of $g$ with respect
to $x$. Using the notation where multiple $x$-derivatives of $g$ are
written as $g_{A_1\cdots A_k}$, the \Ka\ metric $g_{A\bar B}$
coincides with $g_{AB}$, without the need for distinguishing
holomorphic and anti-holomorphic indices,
\begin{equation}
g_{A\bar B} =  6\,\frac{(d\,x)_{AB}}{(d\,xxx)} - 9\, \frac{(d\,
xx)_A\,(d\,xx)_B}{(d\,xxx)^2} .
\label{scalkin}
\end{equation}
The domain for the variables is restricted by the requirement that the
kinetic terms for the scalars and for the graviton have the
appropriate signature. This implies that \eqn{scalkin} is
negative definite and $dxxx>0$.

The derivatives of $g$ are homogeneous functions of $x$, so that
\begin{equation}
x^Ag_{A\Bk}=-k\,g_{\Bk}\ .
\end{equation}
This implies in particular that
\begin{equation}
x^A=-g^{AB}g_B\ .
\end{equation}

The curvature corresponding to (\ref{scalkin}) is
\begin{equation}
R^A{}_{\!BC}{}^D = - 2\delta_{(B}^A\,\delta_{C)}^D +
\textstyle{\frac{4}{3}} C^{ADE} \,d_{BCE} \ ,\label{curvC}
\end{equation}
where $C^{ABC}$ equals
\begin{equation}
C^{ABC} = -27 g^{AD}g^{BE}g^{CF}d_{DEF}(d\,xxx)^{-2}.
\label{defCABC}\end{equation}
This tensor is only constant if we are dealing with a
symmetric space. In that case, the curvature components $R^A{}_{BC}{}^D$
are thus constant (in the special coordinates introduced above)
as well as covariantly constant.
The homogeneity properties imply that
\begin{equation}
x^CR^A{}_{\!BC}{}^{\!D}=-g^{AE}g^{DF}g_{BEF}\ .\label{hompropR}
\end{equation}

We determine all possible isometries of the metric
\eqn{scalkin}. Subsequently the result will be compared to the
duality  invariances of $N=2$ Maxwell-Einstein supergravity in
four  dimensions, found in \cite{BEC}.
For the purpose of the proof we first note the existence of
a subgroup of isometries, generated by constant
real translations with parameters $b^A$ and scale
transformations with parameter $\beta$,
\begin{equation}
\delta z^A = b^A - \textstyle{2\over 3} \beta \,z^A \, ,
\label{rztrans}
\end{equation}
which are present irrespective of the value of the coefficients
$d_{ABC}$. These symmetries can be understood from the
dimensional reduction and find their origin in the symmetries of
the corresponding five-dimensional theory
\cite{dWVPSwa}.

The Killing equations that we must solve are
\begin{eqnarray}
&&g_{DA}\partial _{\bar B}\xi^D +g_{D\bar B}\partial _{A}\xi^D
=0 \,,\label{kil1} \\
&&g_{DB}\partial_A \xi ^D + g_{DA} \partial _{\bar B}\xi ^{\bar D}
+ig_{ABD}(\xi ^D -\xi ^{\bar D})=0 \,.\label{kil2}
\end{eqnarray}
Before turning to these equations we shall motivate a special
parametrization of the Killing vectors $\xi^A$ that we will use
below. First we may
always consider the commutator of any isometry with the
translations and the scale transformations (\ref{rztrans}). As
the result must again be an isometry corresponding to some
Killing vector, we conclude that
under a uniform scaling of the $z^A$ and $\bar z^A$, or
equivalently of the real coordinates $x^A$ and $y^A$, every
Killing vector can again be decomposed into the independent
Killing vectors. A similar result holds with respect to the
translations: the derivative of any Killing vector with respect
to $y^A$ must again be decomposable into Killing
vectors. As the number of independent Killing vectors is
finite, the latter restricts us to polynomials in $y$ of
a certain degree $p$
(and $x$-dependent coefficient functions) and/or to a
finite set of exponential functions. However, the exponential
functions in $y$ are not viable in view of the requirement from
the scale transformations. When expanding in $y$ they require an
infinite set of coefficient functions in $x$ in order to ensure
that scale transformations will not generate an infinite number of
Killing vectors (stated differently, the Killing vectors should
be decomposable into a finite set of homogeneous functions in $x$
and $y$). However, taking multiple derivatives with respect to
$y$ then still leads to an infinite set of Killing vectors. On
the basis of these arguments we conclude that the Killing vectors
can be written as {\em finite} polynomials in the $y^A$,
multiplied by homogeneous coefficient functions of $x^A$ of a
degree such that the full Killing vector is  homogeneous
in $x^A$ and $y^A$. Hence we may write
\begin{equation}
\xi ^A(x,y)=\sum_{k=0}^{p}\xi ^A_{B_1 \cdots B_k}(x) \;
y^{B_1} \cdots y^{B_k} \ . \label{kansatz}
\end{equation}

With this parametrization, we write down the Killing
equations for the terms proportional to $y^{B_1} \cdots y^{B_k}$.
The first Killing equation (\ref{kil1}) implies
\begin{equation}
g_{D(A}\, \left\{\partial_{B)}\xi^D_{B_1\cdots B_k}(x) +i(k+1)\,
\xi^D_{B)B_1\cdots B_K}(x)\right\} =0\ ,  \label{Ki1}
\end{equation}
where here and henceforth the derivatives
act on functions of $x$ only and $\partial _B$ will denote the
derivative with respect to $x^B$ and {\em not} with respect to $z^B$,
as before. The second Killing equation \eqn{kil2} reads
\begin{eqnarray}
&&ig_{DB}\partial _A \xi ^D_{\Bk} -ig_{DA}\partial _B\bar \xi^D_{\Bk}
\label{Ki2} \\
&&+(k+1)\left( g_{DB}\xi^D_{A\Bk} +g_{DA}\bar \xi ^D_{B\Bk}\right)
+i(\xi -\bar \xi )^D_{\Bk} g_{ABD}=0 \ .
\nonumber
\end{eqnarray}
Decomposing the functions $\xi^A_{B_1\cdots B_k}$
into real and imaginary parts, $R^A_{B_1\cdots B_k}$ and
$I^A_{B_1\cdots B_k}$, respectively, the Killing equations
correspond to three real equations,
\begin{eqnarray}
&& \partial_AR^D_{B_1\cdots B_k}  - (k+1)\,g_{AE}\,g^{DF}\,
I^E_{FB_1\cdots B_k}=0 \ ,  \label{een}\\
&& g_{D(A}\,\left(\partial_{B)} I^D_{B_1\cdots B_k} + (k+1)\,
R^D_{B)B_1\cdots B_k}\right) = 0\ ,
\label{twee} \\
&& g_{ABD}\, I^D_{B_1\cdots B_k}  - 2(k+1)g_{D(A}\,
R^D_{B)B_1\cdots B_k}=0 \ .    \label{drie}
\end{eqnarray}

Let us first evaluate (\ref{drie}) a little further. After
multiplication with $x^B$, $I^D_{\Bk} (x)$ is expressed in terms
of $R^D_{B_1\cdots B_{k+1}} (x)$,
\begin{equation}
I^D_{\Bk} =\half (k+1) \left( g_E \,g^{BD} -x^B\,
\delta^D_E\right) R^E_{B\Bk}  \ .\label{imag}
\end{equation}
Re-inserting this result back into \eqn{drie} gives
\begin{equation}
\left( 4\delta ^F_{(A}\,g_{B)E} - g_{ABD} \,g^{DF}\,g_E +g_{ABE}\,
x^F \right)
R^E_{F\Bk}=0\ .  \label{vier}
\end{equation}
By virtue of \eqn{imag} we can express
\eqn{een} and \eqn{twee} exclusively in
terms of the real parts by using \eqn{imag}. The resulting
equations can be written as
\begin{eqnarray}
&&\partial _A R^D_{\Bk} = \half (k+1)(k+2) \left(
g_E \delta ^B_A-g_{AE}x^B\right) g^{DF}R^E_{BF\Bk}\ ,
\label{dReR} \\
&&\half g_{D(A}\partial _{B)}\left( \big(g_E g^{FD}-x^F \delta
_E^D\big) R^E_{F\Bk}\right) +g_{D(A}R^D_{B)\Bk}=0 \ .
\end{eqnarray}
Using \eqn{vier} we can, in the left-hand side of the
last equation, bring $g_{DA}$ under the derivative and obtain
\begin{equation}
\partial _A \partial _B\left( x^Fg_E R^E_{F\Bk}\right) =
2g_D \partial _{(A} R^D_{B)\Bk}
+\partial _{(B}\left(\big(\partial _{A)}R^D_{F\Bk}\big)x^F g_D \right).
\label{eqC}
\end{equation}
At this stage we are thus left with the equations \eqn{vier},
\eqn{dReR} and \eqn{eqC}, while the imaginary parts follow simply
from \eqn{imag}.

Now we concentrate on the leading term in the expansion
\eqn{kansatz}. Choosing $k=p$ in \eqn{een} and
\eqn{imag} shows at once that $\xi^A_{B_1\cdots B_p}$ is real and
constant. Therefore the Killing vector is homogeneous of degree
$p$, so that the functions $\xi^A_{B_1\cdots B_k}$ are homogeneous
of degree $p-k$. Hence it follows from \eqn{dReR} that
\begin{equation}
(p-k) R^D_{B_1\cdots B_k} = (k+1)(k+2) \,
g^{DC}\left(x^B\,g_E\, R^E_{BCB_1\cdots B_k }\right) \ ,\label{Rrec}
\end{equation}
Furthermore, multiplying \eqn{eqC} with $x^A\,x^B$ and using the
homogeneity of the functions involved, one shows at once that
\begin{equation}
x^B\,g_E\,R^E_{BB_1\cdots B_k}= 0 \ \ \  \mbox{for} \ \
k\not=p-1\ .    \label{contr1}
\end{equation}
On the other hand $g_D\,x^C\,R^D_{CB_1\cdots B_{p-1}}$
is homogeneous of zeroth degree (while $R^D_{B_1\cdots B_p}$
itself is constant).  As its second derivative must
vanish according to \eqn{eqC}, it follows that it must be
equal to a constant symmetric tensor,
\begin{equation}
g_D\,x^C\,R^D_{CB_1\cdots B_{p-1}}= C_{B_1\cdots B_{p-1}}\ ,
\label{contr2}
\end{equation}
According to \eqn{Rrec} and \eqn{contr1} only $R^D_{B_1 \cdots
B_p}$ and $R^D_{B_1 \cdots B_{p-2}}$ can now be nonvanishing,
with the former restricted by \eqn{contr2} and the latter given by
\begin{equation}
R^D_{B_1\cdots B_{p-2}} = \half p(p-1) \,
g^{DC} C_{CB_1\cdots B_{p-2}} \ .   \label{Rpm2}
\end{equation}

We now proceed as follows. We analyze the equations
\eqn{contr1}--\eqn{Rpm2}, first for $p>2$, where we find that there are
no non-trivial solutions, and subsequently for $p\leq 2$. Then
we verify whether the solutions satisfy the conditions
\eqn{vier}, \eqn{dReR} and \eqn{eqC}. For the maximal non-trivial
value of $k$ these conditions are already satisfied by virtue of the
results obtained so far. For instance, \eqn{vier} with
$k=p-1$, follows from \eqn{contr2}, as one can verify by taking
its first- and second-order derivative. Hence we are left with
\eqn{vier} for $k=p-3$, \eqn{dReR} for $k=p-2$ and $k=p-4$, and
\eqn{eqC} for $k=p-3$. However, because it turns out that
non-trivial solutions exist only for $p\leq2$, we will only have to
verify the validity of \eqn{dReR} for $p=2$ and $k=0$.

For \underline{$p>2$} we get from \eqn{Rpm2} that
\begin{equation}
g_D\,x^E\,R^D_{EB_1\cdots B_{p-3}} =-\half p(p-1) \, x^A\,x^C\,
 C_{ACB_1\cdots B_{p-3}}\ ,
\end{equation}
which must vanish by virtue of \eqn{contr1}. Therefore we find
that $C_{B_1\cdots B_{p-1}}$ vanishes. For
those values of $p$ we are thus left with only
$R^D_{B_1\cdots B_p}$ satisfying
$g_E\,x^C\,R^E_{CB_1\cdots B_{p-1}}=0$. However, taking the
derivative with respect to $x^A$ and contracting
with $x^{B_1}$ then leads to
\begin{equation}
g_{AE}\,x^C\,x^D\,R^E_{CDB_2\cdots B_{p-1}} =0\ ,
  \label{R0ifC0}
\end{equation}
so that also $R^D_{B_1\cdots B_p}$ vanishes.

For $\underline{p=0}$ we find just the real translations
introduced in \eqn{rztrans} with $R^A=b^A$.

For $\underline{p=1}$ \eqn{contr2} reads
\begin{equation}
x^B\,g_A\, R^A_B = C\ ,  \label{p1cond}
\end{equation}
which, upon differentiation, leads to
\begin{equation}
g_A\,R^A_B + g_{BA}\, R^A_C\,x^C  =0\ .
\end{equation}
Then we obtain $I^A$ from \eqn{imag}
so that
\begin{equation}
\xi^A= 2R^A_B\, z^B \ .
\end{equation}
The condition \eqn{p1cond} can be written as
\begin{equation}
3 d_{E(AB}\,R^E_{C)} = C\, d_{ABC}\ .
\end{equation}
The solution can be decomposed into a particular solution
$R^E_C\propto \delta ^E_C$ and a solution of the homogenous
equation for which the
left-hand side is zero. The latter are the matrices that leave
the tensor $d_{ABC}$ invariant. Hence we can write
$R^A{}_B=
-\ft23 \beta\,\delta^A_B + \tilde B^A{}_{\!B}$, with $\beta $ real
(the normalization is such that we reproduce the scale
transformation in \eqn{rztrans}), while $\tilde B^A{}_{\!B}$ should
satisfy \eqn{Binv}.

For $\underline{p=2}$ the condition \eqn{contr2} reads
\begin{equation}
x^B\,g_E\, R^E_{AB}  = C_A\ ,
\end{equation}
and leads to
\begin{equation}
g_E\,R^E_{AC} + g_{CE}\, R^E_{AB}\,x^B  =0\ .
\end{equation}
If $C_A=0$ we can use the same argument as before (cf.
\eqn{R0ifC0}) to put the whole solution to zero. So we need only
a particular solution of the inhomogeneous equation.  Therefore
we note that \eqn{hompropR} implies the relation
\begin{equation}
g_A\,x^C\,R^A{}_{\!BC}{}^{\!D}= -2\delta _B^D \ ,
\end{equation}
so that we obtain the following solution for the coefficients
$R^E_{AB}$,
\begin{equation}
R^E_{AB}=- \half R^E{}_{\!AB}{}^{\!D} C_D\ .
\end{equation}
In addition we must require that the right-hand side is constant.
Subsequently,
from \eqn{imag} we obtain $I^A_B$, while \eqn{Rpm2} determines
$R^A$, such that
\begin{equation}
\xi^A=4R^A_{BC}\, z^B\,z^C \ .
\end{equation}
One can verify that also \eqn{dReR} is satisfied for this solution, which
completes the analysis.

We have therefore found the following isometries (we redefine
$C_A=-\ft14 a_A$)
\begin{equation}
\delta z^A = b^A - \textstyle{2\over 3} \beta \,z^A +\tilde
B^A_{\;B}\, z^B + \textstyle{1\over 2} \big(R^A{}_{\!BC}{}^{\!D}\,
a_D\big)\, z^B z^C .
\label{ztrans}
\end{equation}
where the parameters are real,
$\tilde B$ satisfies \eqn{Binv}, and $R^E{}_{\!AB}{}^{\!D} a_D$
should be constant.
These are exactly the generalized duality transformations
found in \cite{BEC}.

There is one point that needs further clarification.
In \cite{BEC} the condition on the parameters $a_A$ was
given as
\begin{equation}
a_{G}\, E^G_{ABCD}  = 0\ ,  \label{ainv}
\end{equation}
where the tensor $E^G_{ABCD}$ is defined by
\begin{equation}
E^G_{ABCD} = C^{EFG}\, d_{E(AB}\,d_{CD)F} -  \delta^G_{(A} \,
d_{BCD)}\ ,
\label{defE}
\end{equation}
and $C^{ABC}$ was given in \eqn{defCABC}.
To see that \eqn{ainv} is equivalent to the condition given
above, we note the relations
\begin{equation}
{\textstyle \frac{1}{18}}(dxxx)^2\,g_{AE}\,
g_{BF}\,g_{CG}\,
\frac{\partial }{\partial x^D} C^{EFG}=-6g_{DF}\,E^F_{ABCE}\,x^E
=6g_E\, E^E_{ABCD}\ .   \label{relCCE}
\end{equation}
They allow us to derive the following equivalence
\begin{equation}
E^E_{ABCD}a_E=0 \,\Longleftrightarrow
\, \frac{\partial }{\partial x^D}C^{ABC}a_C=0 \,\Longleftrightarrow
\, \frac{\partial }{\partial x^F} R^E{}_{\!AB}{}^{\!D} a_D =0\ .
\label{4conda}
\end{equation}
The second and third condition are equivalent by virtue
of \eqn{curvC}. The first condition implies the second one, as
follows directly from \eqn{relCCE}. Conversely, the second
condition implies that $a_E E^E_{ABCD}x^D=0$, again
because of \eqn{relCCE}. As $C^{ABC}a_C$ is now constant, also
$a_E E^E_{ABCD}$ is constant. Since we know that this constant
should vanish upon multiplication with $x^D$, it should itself be
zero. This completes the proof that all isometries of the special
\Ka\ manifolds based on \eqn{Fd4}
coincide with the generalized duality
invariances of \cite{BEC} and
are invariances of the full supergravity theory.

\vspace{1cm}
\noindent
We thank Fran\c{c}ois Vanderseypen, who contributed to this work
at an early stage. A.V.P. thanks the theory
division of CERN, where part of this work was performed, for
hospitality.

\end{document}